\documentclass[useAMS,usenatbib]{mn2e}

\def \nh {$N_{H}$}

\def\ergs {erg\,s$^{-1}$}
\def\ergscm2 {erg\,s$^{-1}$cm$^{-2}$}

\def\uu {4U\,0142+614\,}
\def\ee {1E\,1048.1--5937\,}

\def\rxj {1RXS\,J170849.0--400910\,}
\def\ea {1E\,2259+586\,}

\newcommand{\XMM}{{\it XMM--Newton}\,}
\newcommand{\BSAX}{{\it Beppo}SAX\,}
\newcommand{\RXTE}{{\it R}XTE\,}
\newcommand{\INT}{{\it INTEGRAL}\,}
\newcommand{\CXO}{{\it Chandra}\,}

\input psfig.sty

\title[Very deep XMM--Newton observations of \uu\,]{Very deep X-ray observations of the Anomalous X-ray Pulsar 4U\,0142+614}
\author[Rea et al.]{N. Rea$^{1}$\thanks{N.Rea@sron.nl}, E. Nichelli$^{2,3}$,  G.L. Israel$^{2}$, R. Perna$^{4}$, T. Oosterbroek$^{5}$, A.~N. Parmar$^{6}$, 
\newauthor R. Turolla$^{7}$, S. Campana$^{8}$, L. Stella$^{2}$, S. Zane$^{9}$, L. Angelini$^{10}$ \\
$^{1}$SRON - Netherlands Institute for Space Research, Sorbonnelaan 2, 3584 CA, Utrecht, The Netherlands \\
$^{2}$INAF--Astronomical Observatory of Rome, Via Frascati 33, 00040 Monteporzio Catone (Rome), Italy \\
$^{3}$Dublin Institute for Advanced Studies, 5 Merrion Square, Dublin 2, Ireland \\
$^{4}$JILA and Department of Astrophysical and Planetary Sciences, University of Colorado, 440 UCB, Boulder, 80309, USA \\
$^{5}$Science Payload and Advanced Concepts Office, ESA, ESTEC, Postbus 299, 2200 AG, Noordwijk, The Netherlands \\
$^{6}$Research and Scientific Support Department of ESA, ESTEC, Postbus 299, 2200 AG Noordwijk, The Netherlands \\
$^{7}$University of Padua, Physics Department, via Marzolo 8, 35131, Padova, Italy \\
$^{8}$INAF--Astronomical Observatory of Brera, via Bianchi 46, 23807 Merate (Lc), Italy \\
$^{9}$Mullard Space Science Laboratory, University College London, Holmbury St. Mary, Dorking Surrey, RH5 6NT, UK \\
$^{10}$NASA, Goddard Space Flight Center, Greenbelt, MD, USA}

\begin{document}

\date{Submitted on 28/02/07; Accepted on 23/07/07}

\pagerange{\pageref{firstpage}--\pageref{lastpage}} \pubyear{2002}

\maketitle

\label{firstpage}

\begin{abstract}

  We report on two new \XMM\, observations of the Anomalous X-ray
  Pulsar (AXP) \uu\, performed in March and July 2004, collecting the
  most accurate spectrum for this source. Furthermore, we analyse two
  short archival observations performed in February 2002 and January
  2003 (the latter already reported by G\"ohler et al. 2005) in order
  to study the long term behaviour of this AXP. \uu\, appears to be
  relatively steady in flux between 2002 and 2004, and the
  phase-averaged spectrum does not show any significant variability
  between the four epochs. We derive the deepest upper limits to date
  on the presence of lines in \uu\, spectrum as a function of energy:
  equivalent width in the 1--3\,keV energy range $<$~4\,eV and
  $<$~8\,eV for narrow and broad lines, respectively. A remarkable
  energy dependence in both the pulse profile and the pulsed fraction
  is detected, and consequently pulse-phase spectroscopy shows
  spectral variability as a function of phase. By making use of \XMM\
  and \INT\ data, we successfully model the 1--250\,keV spectrum of
  \uu\, with three models presented in Rea et al. (2007a), namely the
  canonical absorbed blackbody plus two power--laws, a resonant
  cyclotron scattering model plus one power--law and two log-parabolic
  functions.

\end{abstract}

\begin{keywords}
stars: pulsars: individual: \uu\, -- stars: magnetic fields -- stars: neutron -- X-rays: stars
\end{keywords}

\section{Introduction}

Anomalous X-ray Pulsars (AXPs) form a restricted family of neutron
star (NS) X-ray sources with properties quite different from those of
other known classes of X-ray emitting NSs, in particular the radio
pulsars. However, in the last few years evidence gathered of a
possible link between AXPs and another peculiar class of NSs: the Soft
Gamma--ray Repeaters (SGRs; see Woods \& Thompson 2006 for a recent
review).

The nature of the strong X-ray emission from AXPs and SGRs is
intriguing. In fact, their X-ray luminosity is too high to be powered
by the loss of the star rotational energy alone, as in the more common
radio pulsars. At the same time, the lack of observational signatures
of a companion strongly argues against an accretion powered binary
system, favouring instead scenarios involving isolated NSs.


\begin{figure*}
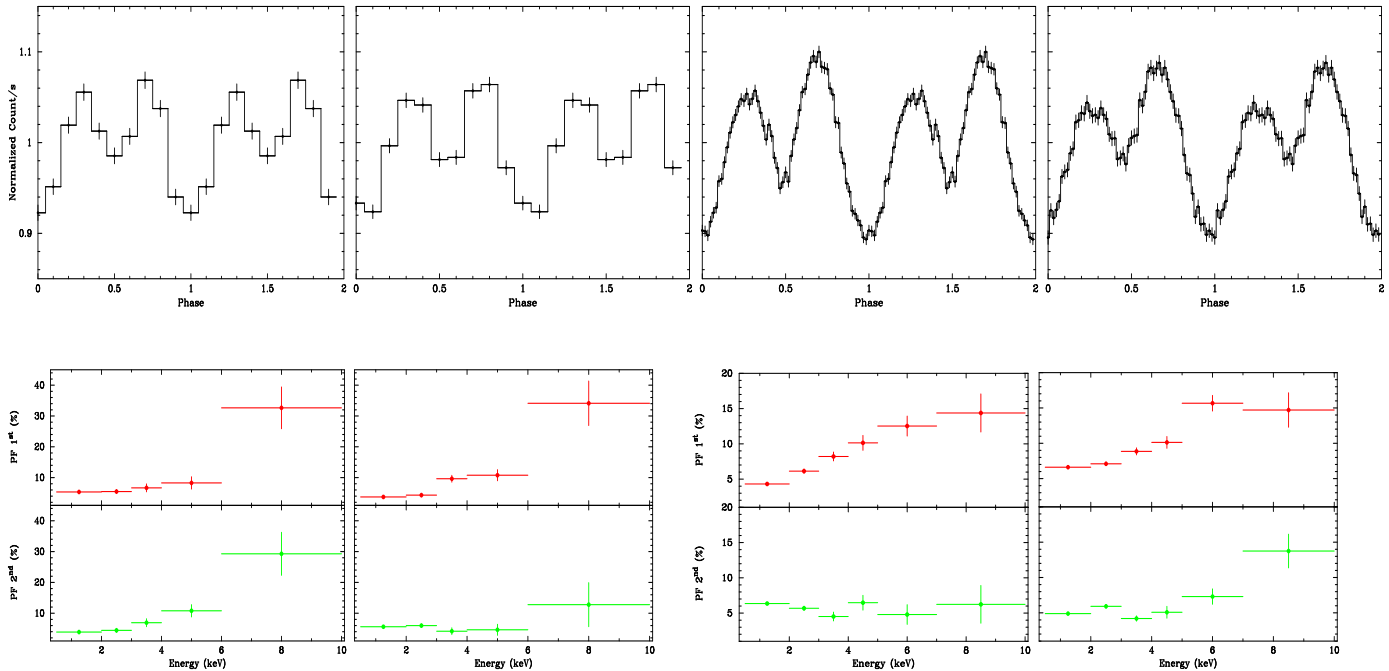

\centerline{
\vbox{
\hbox{
\psfig{figure=efold_final_2002.ps,width=4.5cm,height=4cm,angle=-90}
\psfig{figure=efold_final_2003.ps,width=4.5cm,height=4cm,angle=-90}
\psfig{figure=efold_final_march2.ps,width=4.5cm,height=4cm,angle=-90}
\psfig{figure=efold_final_july2.ps,width=4.5cm,height=4cm,angle=-90} }
\vspace{0.8cm}
\hbox{ 
\psfig{figure=pf_vs_energy_2002_c.ps,width=4.5cm,height=4cm,angle=-90}
\psfig{figure=pf_vs_energy_2003_2.ps,width=4cm,height=4cm,angle=-90}
\hspace{0.3cm}
\psfig{figure=pf_vs_energy_march_c.ps,width=4.5cm,height=4cm,angle=-90}
\psfig{figure=pf_vs_energy_july_2.ps,width=4cm,height=3.96cm,angle=-90} } } }

\caption{{\em Top row}: 0.3--10\,keV background-subtracted pulse
    profiles of all the four \XMM\, observations of \uu. {\em Bottom
    row}: pulsed fraction as a function of energy for all the four
    \XMM\, observations of \uu: the top and bottom panels report,
    respectively, on the pulsed fraction (as defined in the text) of
    the fundamental and first harmonic sine functions which best fit
    the pulse profile of each observation (note that first two and
    last two panels from left have different y-axis units). For both
    top and bottom rows, the relative observations are (from left to
    right): February 2002, January 2003, March 2004 and July 2004.}

\label{pfene}
\label{fig:powspec}
\end{figure*}


AXPs are usually radio-quiet and exhibit X-ray pulsations with spin
periods in the $\sim 5$--12\,s range. This, along with their large
spin-down rates ($\dot{P}\approx 10^{-13}-10^{-10}$s\,s$^{-1})$,
implies magnetic fields exceeding the (electron) critical magnetic
field, $B_{\rm cr}\sim 4.4\times10^{13}$~G, if spin-down occurs mainly
through magneto-dipolar braking.

Furthermore, they are characterised by a high X-ray luminosity
($L_{X}\approx 10^{34}$--$10^{36}$\ergs). Their high-energy emission
($\sim 0.5$--100 keV) is traditionally modelled by an absorbed
blackbody ($kT\sim0.4$\,keV) and two power--laws ($\Gamma_{\rm
soft}\sim 3$ and $\Gamma_{\rm hard}\sim 1$), although recently several
attempts have been made to apply more physical spectral models
(e.g. Perna et al. 2001; Rea et al.~2007a, b, c).

At present the model which appears most successful in explaining the
peculiar observational properties of AXPs and SGRs is the ``magnetar''
model. In this scenario AXPs and SGRs are thought to be isolated NSs
endowed with ultra-high magnetic fields ($B\sim 10^{14}-10^{15}$\,G)
the decay of which powers their X-ray emission (Duncan \& Thompson
1992; Thompson \& Duncan 1993, 1995, 1996). At magnetic fields higher
than $B_{\rm cr}$, radio emission is believed to be suppressed e.g. by
the photon--splitting process (Baring \& Harding 1998), and this could
explain why magnetars were believed not to show any radio
emission. However, the discovery of radio pulsars with magnetic fields
higher than the electron critical field (Camilo et al.~2000) and with
very weak X--ray emission, as well as the discovery of radio
pulsations from one magnetar (Camilo et al.~2006), is puzzling and
still awaits explanation. Alternative scenarios, invoking accretion
from a fossil disk, remnant of the supernova explosion (van Paradijs,
Taam \& van den Heuvel 1995; Chatterjee, Hernquist \& Narayan 2000;
Perna et al.~2000; Alpar 2001), encounter increasing difficulties in
explaining the data.

\uu\, is one of the brightest AXPs known to date, and it was first
detected by {\it Uhuru} in 1978 (Forman et al. 1978). However, mainly
because of the presence of the accretion-powered binary pulsar RX
J0146.9+6121 nearby, only in 1994 was an $\sim8$\,s periodicity
reported using EXOSAT data taken in 1984 (Israel, Mereghetti \&
Stella~1994).  Long term spin period variations were discovered thanks
to a large \RXTE\, campaign (Gavriil \& Kaspi 2002), leading to the
measure of the period derivative $\dot{P} \sim 2 \times
10^{-12}$\,s\,s$^{-1}$.  Despite deep searches (Israel, Mereghetti \&
Stella 1994, Wilson et al. 1999), no evidence for orbital motion has
been found, supporting the isolated NS scenario.

Subsequent observations (White et al. 1996; Israel et al. 1999;
Paul et al.  2000) revealed a soft (1--10\,keV) X--ray spectrum
typical of an AXP, best fitted by an absorbed blackbody ($kT \sim
0.4$\,keV) plus a power law ($\Gamma \sim 3.7$). More recent {\it
Chandra} (Juett et al. 2002; Patel et al. 2003), \XMM\, (G\"ohler
et al.~2005) and {\it Swift--XRT} (Rea et al.~2007a) observations
have shown that \uu\, is a relatively stable X--ray emitter.


\begin{figure*}
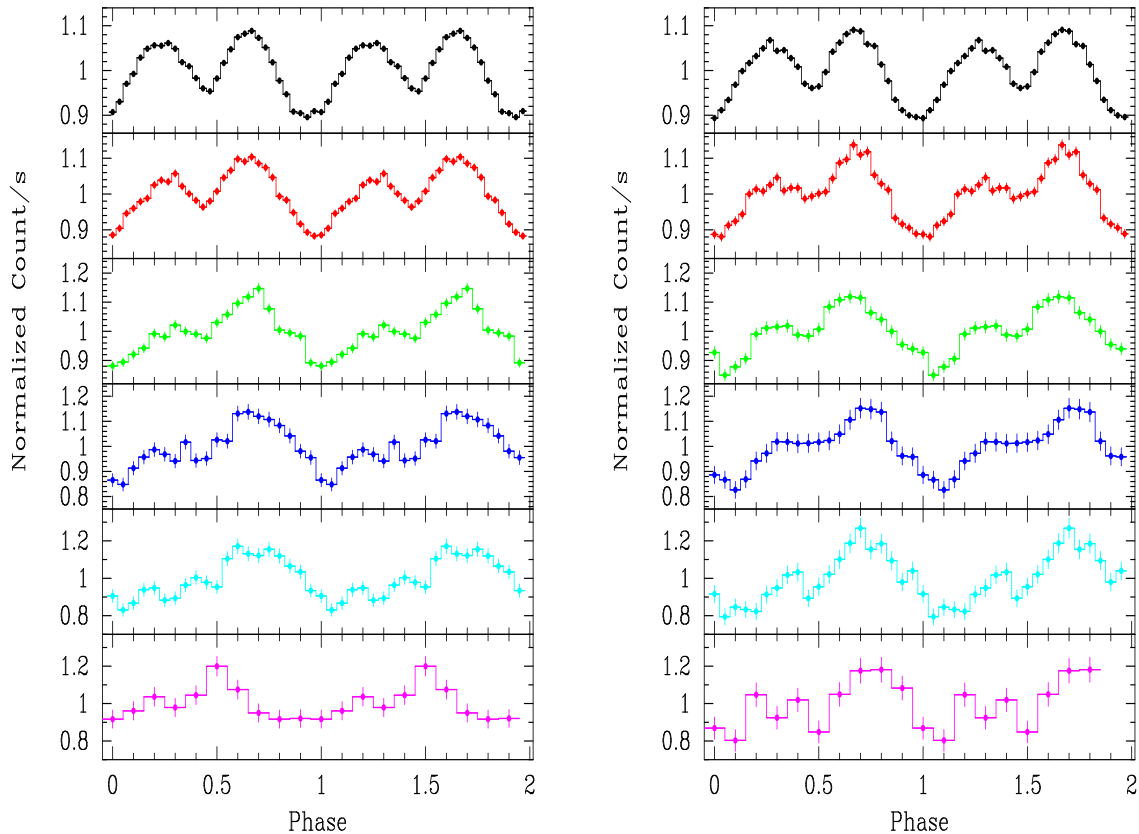

\centerline{
\hbox{
\psfig{figure=march_efold_energy_c.ps,height=11cm,width=7cm,angle=-90}
\hspace{0.8cm}
\psfig{figure=july_efold_energy_c.ps,height=11cm,width=7cm,angle=-90}}  }
\caption{Epoch folding as a function of energy (from top to bottom: 0.5--2 keV,
2--3 keV, 3--4 keV, 4--5 keV, 5--7 keV, 7--10 keV) for the 2004 March (left
panel) and July (right panel) observations.}
\label{efoldenergy}
\end{figure*}

In the last few years, two peculiar characteristics of \uu, have been
found in comparison with other AXPs: {\em i)} an optical counterpart
(Hulleman, van Kerlwijk \& Kulkarni 2000), displaying $\sim8$\,s
pulsation with a $30\%$ pulsed fraction (Kern \& Martin
2001)\footnote{Note that the detection of the optical counterpart to
this object might be due to its lower extinction with respect to other
AXPs, rather than to an intrinsic peculiarity.} and {\em ii)}
mid--infrared emission, tentatively interpreted as the signature of a
non--accreting disk around the NS (Wang, Chakrabarti \&
Kaplan~2006). Furthermore, like in other AXPs, a hard X--ray emission
up to 250\,keV has been revealed (Kuiper et al.~2006; den~Hartog et
al.~2006, 2007).

In this paper we report on two new \XMM\, observations of the AXP
\uu\, performed in March and July 2004. These observations provide by far
the most accurate data collected in the soft X-ray band for this
source. Furthermore, in order to study the long term X--ray
activity of \uu\, we analysed two archival observations (one of
them already published by G\"ohler et al.~2005). Data analysis and
results are presented in
\S~\ref{obs}, while \S~\ref{multiband} deals with the spectral
modelling of the 1--250\,keV spectrum of
\uu. Finally our results are discussed in \S\,\ref{discussion}.

\section{Observations, Data Analysis and Results}
\label{obs}

\uu\, has been observed with \XMM\, four times, the first two
times for a very short exposure time ($\sim$6\,ks), while the last two
observations are much deeper. These two very deep observations have
been performed on March 1st and July 25th 2004, with on--source
exposure times of 46 and 23\,ks, respectively (see the end of this
section for further details on the first two short observations).  The
\XMM\, Observatory (Jansen et al. 2001) includes three 1500~cm$^2$
X-ray telescopes with an EPIC instrument in each focus, a Reflecting
Grating Spectrometer (RGS; den Herder et al. 2001) and an Optical
Monitor (Mason et al. 2001). Two of the EPIC imaging spectrometers use
MOS CCDs (Turner et al. 2001) and one uses a pn CCD (Str\"uder et
al. 2001).

Data have been processed using SAS version 7.0.0, and we have employed
the most updated calibration files (CCF) available at the time the
reduction was performed (November 2006). Standard data screening
criteria are applied in the extraction of scientific products. We have
cleaned the March 2004 observation for solar flares (the observation
in July is not affected); the exposure time after the solar flare
filtering is 37\,ks.

Both 2004 observations have the same instrument set--up. pn and MOS1
cameras are used with the {\tt medium} filters, while the {\tt thick}
filter is applied for the MOS2 camera. The pn camera is set in {\tt
timing} mode in order to avoid pile--up, the MOS1 with the central
CCD in {\tt timing} mode, and the others in {\tt prime full window}
mode. The MOS2 CCDs are all set in {\tt prime full window} mode, with
the aim of unveiling possible transient X--ray sources which might
contaminate the instruments observing in {\tt timing} mode. No
transients are present in the MOS2 image. The source is, as expected,
highly affected by pile--up. Hence, we do not consider the MOS2 data
any further.

In order to extract more than 90\% of the source counts, we have 
constructed, for the data obtained with the pn and the MOS1 central
CCD, a one--dimensional image and we fit a Gaussian to the 1D photon
distribution. We have then extracted the source photons from a rectangular
region with RAWX and RAWY coordinates 37,100.75 and 18,197.5. The
background is obtained from a rectangular region of the same size, as
far as possible from the source, on the east side. Only photons with
PATTERN$<=4$ are used for the pn and with PATTERN$<=12$ for the MOS1.


\begin{table*}
\begin{center}
\begin{tabular}{lccccc}
\hline
\hline
 \multicolumn{1}{l}{Parameters} & \multicolumn{1}{c}{February 2002} &
 \multicolumn{1}{c}{January 2003} & \multicolumn{1}{c}{March 2004} &
 \multicolumn{1}{c}{July 2004} & \multicolumn{1}{c}{Joint fit} \\
\hline
$N_{H}$    & $1.0^{+0.1}_{-0.1}$ & $1.1^{+0.1}_{-0.1}$ & $1.01^{+0.02}_{-0.04}$ &  $1.00^{+0.06}_{-0.01}$ & $1.00^{+0.01}_{-0.01}$ \\
 & & & & &  \\
kT\,(keV) & $ 0.40^{+0.05}_{-0.03}$ & $ 0.41^{+0.05}_{-0.05}$ & $0.409^{+0.007}_{-0.001}$ & $0.421^{+0.007}_{-0.004}$ & $0.410^{+0.004}_{-0.002}$ \\
BB Radius (km) &  $ 4.2^{+0.7}_{-0.5}$ & $ 4.4^{+0.8}_{-0.5}$ & $ 4.75^{+0.08}_{-0.06}$  & $4.17^{+0.09}_{-0.07}$ & $4.22^{+0.07}_{-0.02}$  \\
BB flux (\%)  &  $18^{+3}_{-1}$ &  $21^{+2}_{-2}$ & $ 23.8^{+1.8}_{-0.2}$  & $25.2^{+0.2}_{-0.1}$ & $23.5^{+0.1}_{-0.1}$   \\
 & &  & & & \\
$\Gamma$      &  $3.8^{+0.1}_{-0.1}$ & $4.0^{+0.1}_{-0.2}$ & $3.77^{+0.08}_{-0.02}$ & $3.92^{+0.02}_{-0.02}$  &  $3.88^{+0.01}_{-0.01}$ \\  
& &  & & & \\
Total Flux 0.5--10\,keV     &  $1.2^{+0.2}_{-0.1}$ & $1.1^{+0.1}_{-0.1}$ & $1.16^{+0.02}_{-0.01}$  &  $1.18^{+0.02}_{-0.02}$ & $1.18^{+0.01}_{-0.01}$ \\
Total Flux 2--10\,keV        &  $0.6^{+0.3}_{-0.4}$ & $0.6^{+0.3}_{-0.3}$  & $0.58^{+0.07}_{-0.06}$  & $0.59^{+0.02}_{-0.03}$ & $0.58^{+0.01}_{-0.01}$ \\
 & &  & & & \\
$\chi_{\nu}^2$ (d.o.f.)     & 0.96 (88) &  0.98 (90) & 1.04 (210) &  1.00 (210) & 1.01 (590) \\

\hline
\hline
\end{tabular}
\caption{Parameters of the spectral modelling of the \uu\, spectra 
with an absorbed blackbody plus a power--law, for all the four \XMM\,
observations. The last column reports on the parameters derived from a
joint fit of all four spectra with the latter model. Uncertainties in
this table are given at 90\% confidence level. $N_{H}$ is in units of
$10^{22}$\,cm$^{-2}$ and solar abundances from Anders \& Grevesse
(1989) are assumed. The blackbody radius is calculated assuming a
distance of 3\,kpc (note that in the error calculation we did not
consider the error in the distance). The blackbody flux fraction is
calculated with respect to the absorbed 0.5--10\,keV total
flux. Fluxes are all absorbed and in units of
$10^{-10}$\,erg\,cm$^{-2}$\,s$^{-1}$.}

\end{center}
\end{table*}


From the same regions we have extracted the \uu\, spectra. All of the
EPIC spectra have been rebinned before fitting, so as to have at least
40 counts per bin and not to oversample the energy resolution by more
than a factor three. We have also extracted first and second order
RGS1 and RGS2 spectra for both source and background, using the
standard procedure reported in the \XMM\, analysis manual.

Thanks to the high timing and spectral resolution\footnote{see
http://xmm.esac.esa.int/ for details.} of the pn (0.03\,ms and
E/$\delta$E $\sim 50$) and MOS cameras (1.5\,ms and E/$\delta$E $\sim
50$), and to the high spectroscopical accuracy of the RGS (E/$\delta$E
$\sim 400 $), we are able to perform timing and spectral analysis, as
well as pulse-phase spectroscopy. The RGS spectra do not show evidence
for narrow spectral features, and its continuum spectra are in good
agreement with those of the pn and MOS1. We then do not report further
on RGS data and refer to Durant \& van~Kerkwijk (2006a,b) for details
on those. Furthermore, both MOS1 and pn give consistent timing and
spectral results. Hence, we report hereafter only on the pn data. We
find a pn (background subtracted) count rate of $46.2\pm0.1$ and
$46.5\pm0.1$ count/s for the March and July 2004 observation,
respectively.

Furthermore, we have also analysed two short (6\,ks exposure time
each) \XMM\, observations performed on 2002 February 13th and 2003
January 24th (the latter already reported in detail by G\"ohler et
al.~2005). In both observations the pn camera is set-up in Small
Window mode. We use only pn data for these observations, and refer to
G\"ohler et al.~(2005) for details on MOS data for the January 2003
observation (MOS cameras gave results consistent with the pn for both
observations). We have reprocessed and cleaned the data as reported
above for the 2004 observations, resulting in an on-source exposure
time in 2002 and 2003 of 2.0 and 3.5\,ks, respectively. Thanks to the
2D image provided by the pn in Small Window mode, we have extracted
the source events and spectra from a circular region of
25$^{\prime\prime}$ around the \uu\, position. The background events
and spectra, for both observations, are extracted from a similar
circular region as for the source, although as far as possible (still
within the central CCD) from the \uu\, position. The source pn count
rate is $42.1\pm0.5$ and $40.5\pm0.6$ count/s for the February 2002
and January 2003 observations, respectively.

\subsection{Timing analysis}
\label{timing}

In order to determine the spin period of \uu\,, we have first
corrected the event times to the barycenter of the Solar System (using
the SAS tool {\tt barycen}) and computed a power spectrum ({\tt
powspec}), where we find a strong signal around 8.6\,s in all the
observations.  We use for the timing analysis only events in the
0.3--10\,keV energy range. Timing analysis is performed using {\tt
Xronos}. Two harmonics of the spin period are significantly detected
in all four observations.

A precise estimate of the pulse period in the observations is obtained
carrying out a period search ({\tt efsearch}) around the 8.6\,s
value. We have then refined the period determination by dividing each
observation into six intervals, and performing a linear fit to the
phase determined in each interval (phase-fitting technique).

We obtain a best spin period of $P_{s}$=8.6887(3)\,s (MJD 52318.0),
$P_{s}$=8.6882(2)\,s (MJD 52663.0), $P_{s}$=8.68856(1)\,s (MJD
53065.0) and $P_{s}$=8.688575(2)\,s (MJD 53211.0), for the \XMM\,
observations from February 2002 to July 2004\footnote{All errors in
the text are reported at 1$\sigma$ confidence level, if not otherwise
specified.}. All these periods are consistent within their 3$\sigma$
errors with the more accurate RXTE timing analysis of this source (Dib
et al.~2007).

The pulse profile of \uu\, is double peaked in all observations, and
well modelled by two sine functions with periods fixed at the spin
period and its first harmonic (see Fig.\,\ref{fig:powspec}). Although
the X-ray emission of \uu\, above 100~keV is completely pulsed (pulsed
fraction of $\sim 100$\%; Kuiper et al.~2006), in the soft X-rays
\uu\, shows the smallest pulsed fraction of the class (a factor of
$\sim 3$ below that observed in other AXPs).

The pulsed fraction (defined as the semi--amplitude of the
best-fitting fundamental sine function divided by the, background
corrected, constant level of the emission) in the 0.3--10\,keV energy
band is $5.3\pm0.4$\%, $5.9\pm0.4$\%, $4.7\pm0.2$\% and $5.4\pm0.2$\%,
from February 2002 to July 2004. For a better comparison with previous
results, we report also the peak-to-peak 0.3--10\,keV pulsed fraction
(defined as $(F_{max} - F_{min})/(F_{min} + F_{max})$), again from
2002 to 2004: $8.8\pm0.9$\%, $7.7\pm0.9$\%, $10.2\pm0.2$\% and
$13.3\pm0.3$\%.

The 0.3--10\,keV peak-to-peak pulsed fraction shows a modest ($\sim
2.5\sigma$) increase between 2002 and 2004, possibly consistent with
the evidence for a slow increase of the pulsed flux observed by
\RXTE\, (Dib et al. 2007).  However, this is not followed by a similar
increase of the fundamental sine-function pulsed fraction, which
instead is within the errors relatively constant in time. This might
be explained with the small changes of the pulse profile due mainly to
the first harmonic of the signal, rather than to the fundamental.


\begin{figure*}
\centerline{
\hbox{
\psfig{figure=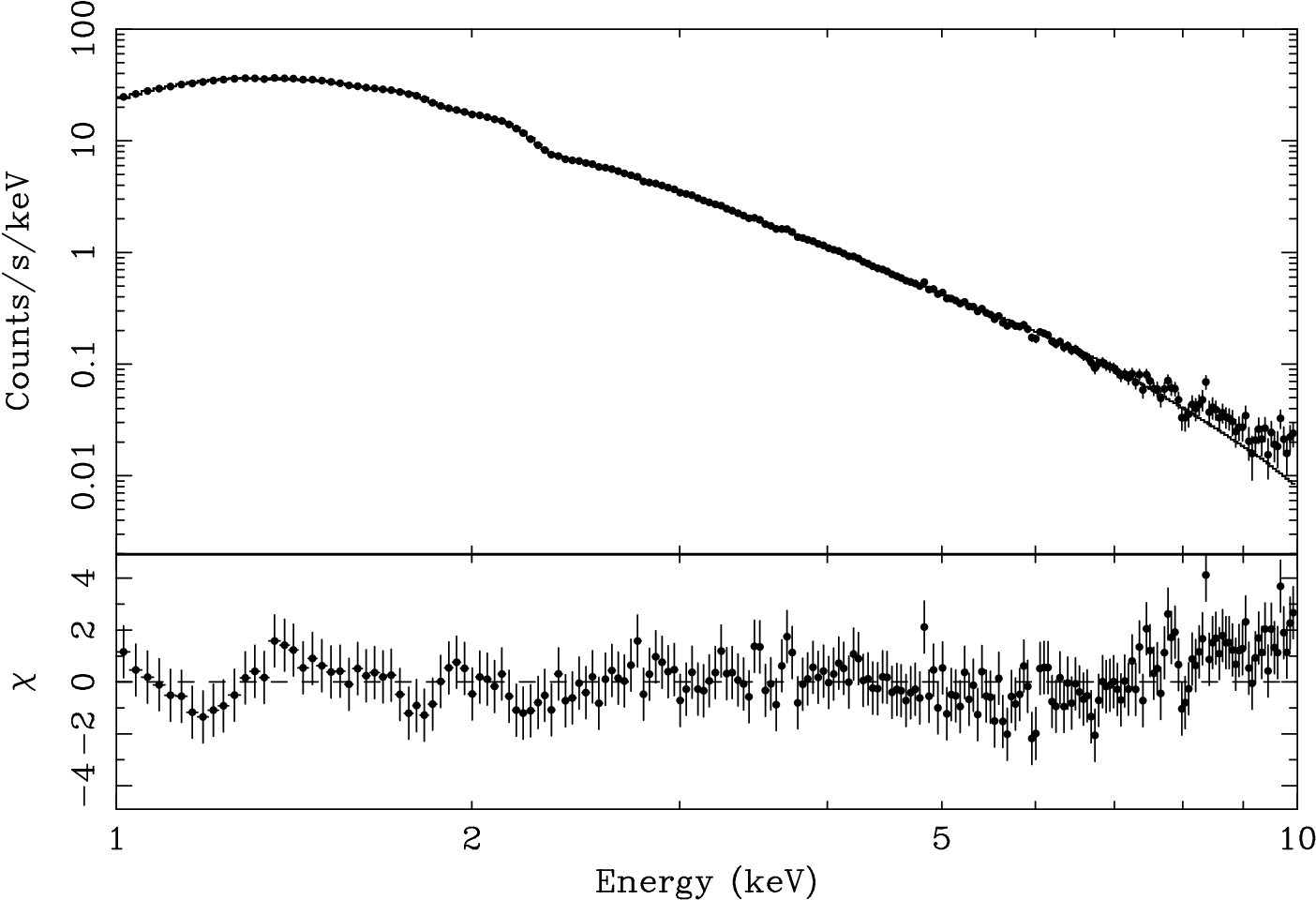,width=7cm,height=5.5cm,angle=-90}
\hspace{1.0cm}
\psfig{figure=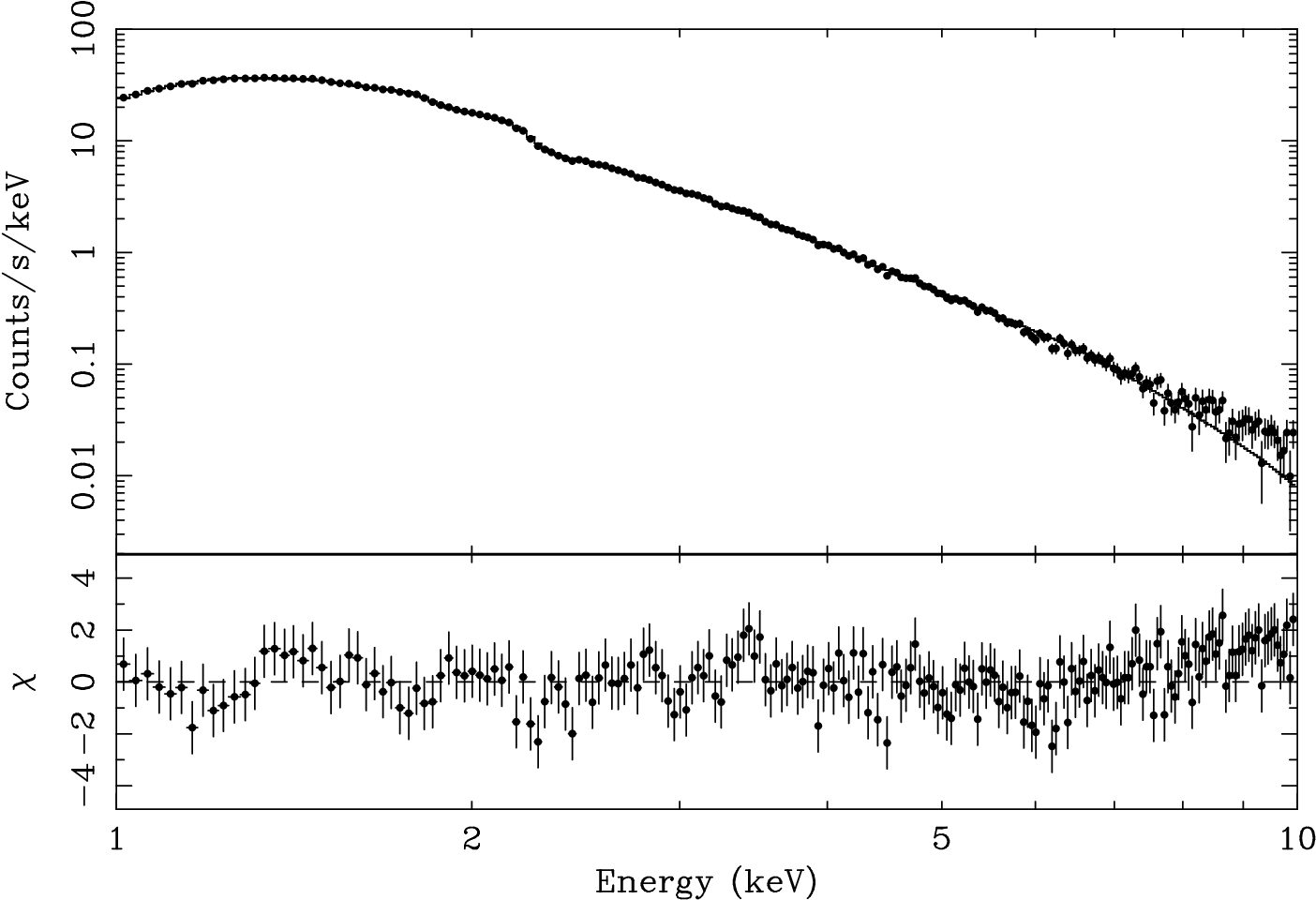,width=7cm,height=5.5cm,angle=-90}}  }
\caption{Phase average spectra of the March (left) and July (right) \XMM\,
observations of \uu\, fitted with an absorbed blackbody plus a power--law model.}
\label{fig:spectra}
\end{figure*}


Studying in detail the pulsed fraction as a function of energy, in all
the observations, we find a peculiar pulsed fraction increase with
energy: e.g. in March and July 2004 the fundamental harmonic pulsed
fraction increases from about 4\% to 15\% as the energy increases from
1 to 10\,keV (see Fig.\,\ref{pfene}). Given the large errors on the
pulsed fraction values in the 6-10\,keV range for the February 2002
and January 2003 observations, this increase of the pulsed fraction
with energy is consistent with having the same trend (within the
errors) in all four observations.  This effect was already hinted in
other observations but the lower number of counts collected was not
sufficient to reliably asses it (see e.g. Fig. 5 of Israel et
al. 1999, Paul et al.~2000, and G\"ohler et al.~2005). It is worth to
note that this trend can not be an instrumental effect of the pn
camera since the observations we report are taken in two different pn
modes (Small Window and Timing), and on the other hand this pulsed
fraction increase was not observed in other AXPs or pulsars observed
with the pn. Furthermore the presence of a hint of this increase also
in \BSAX\, (Israel et al. 1999) and ASCA data (Paul et al.~2000) rules
out any possible instrumental effect.

Correspondingly, the pulse profiles are highly variable with energy,
as shown in Fig.\,\ref{efoldenergy}. This variability of the pulse
profile with energy has always been observed in this source (Israel et
al. 1999; Paul et al.~2000; Patel et al. 2003; Morii et al. 2005), as
well as in other AXPs (see e.g. Rea et al.~2005; Mereghetti et
al.~2004). To asses the statistical significance of these pulse
profile variations, we have performed a Kolmogorov--Smirnov test on
the energy resolved pulse profile for the March and July 2004
observations (see Fig.\,\ref{efoldenergy}), and in both observations
the difference between the soft (0.5--2\,keV) and hard (7--10\,keV)
pulse profile is highly significant, giving a probability that the
profiles have the same shape smaller than $10^{-6}$. We have done the
same test between the two observations to asses the pulse profile
variability in time for a given energy range: in all the energy ranges
the probability of the March and July 2004 profiles having the same
shame is $\sim0.02$, hence not very significant. However, note that
RXTE monitoring over the last ten years did detect pulse profile
variability with time, thanks to its higher sensitivity for pulse
profile changes (Dib et al.~2007).
  
This increase with energy of the pulsed fraction of the fundamental
component has not been observed so far in other AXPs, while the pulse
profile change with energy is a more common behaviour (see e.g. Rea et
al.~2005; Mereghetti et al.~2004; Woods et al.~2004).

\subsection{Spectral analysis and pulse--phase spectroscopy}
\label{spectra}

The phase-averaged spectra for all the observations are satisfactorily
fitted by an empirical model consisting of an absorbed blackbody plus
a power law ($\chi_{dof}^2 \simeq1.0$; see Tab.\,1 and
Fig.\,\ref{fig:spectra}). We use the {\tt XSPEC} package (version 11.3
and 12.1) for all fittings, and used the {\tt phabs} absorption model
with the Anders \& Grevesse (1989) solar abundances. A 2\% systematic
error is included in all fits, as well as in the error calculation, to
account for calibration uncertainties. Similarly to what has been done
in the past for other AXPs, we have tried to fit the spectra of \uu\,
with two absorbed blackbodies and with a Comptonized blackbody, but
both trials resulted in a worse fit ($\chi_{dof}^2 > 2.4$). In all the
observations the hydrogen column density is $ N_{H}\simeq1\times
10^{22}$\,cm$^{-2}$, and the absorbed flux in the 0.5--10\,keV range
is $\sim 1.1\times 10^{-10}$\,erg\,cm$^{-2}$\,s$^{-1}$, corresponding
to a luminosity of $\sim1\times10^{35}$\,erg\,s$^{-1}$ (for a 3 kpc
distance). In the 0.5--10 keV band, the blackbody component accounts
for $\sim$\,24\,\% of the total absorbed flux. The blackbody radius,
as derived from the fits, is smaller than the NS size. If the
blackbody emission originates from the star surface, and is then
partially up-scattered to form the power--law component as predicted
by the twisted magnetosphere model (see \S\ref{multiband} and
\ref{discussion}), this would imply that only a fraction of the
surface is emitting.  \footnote{Note that the measure of the emitting
region can be strongly affected by the choice of the spectral
model. For instance, in the case of cooling neutron stars, accounting
for atmospheric effects would result in an emitting area larger than
that inferred assuming blackbody emission (Perna et al. 2001). These
atmospheric models are not direct applicable to AXPs, which are
unlikely to be passive coolers. Nevertheless, the uncertainty on the
spectral shape for the thermal component affect the determination of
the emitting size.}.


\begin{figure}
\centerline{\psfig{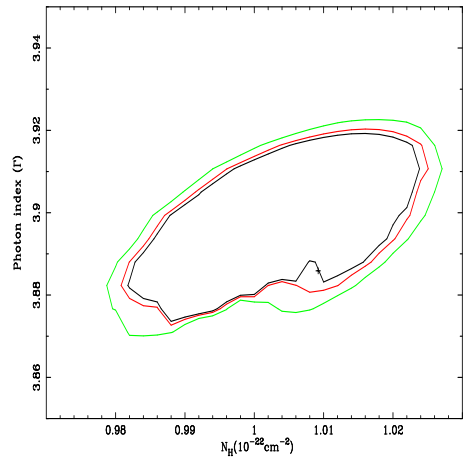}}
\caption{Contours plot of the photon index ($\Gamma$) and the absorption parameter 
for the joint fit reported in Tab.\,1. Confidence levels are from
small to large: 1$\sigma$, 90\% and 99\%. The small cross labels the
values of the joint fit parameters reported in Tab.\,1.}
\label{cntrpl}
\end{figure}


The best fitting parameters for the absorbed blackbody plus power law
model are only marginally ($< 2\sigma$) variable between the
observations (see Tab.\,1; especially between the March and July 2004
pointings), and are all consistent with previous measurements (Israel
et al.~1999; Paul et al.~2000; Juett et al.~2002; Patel et
al.~2003). In order to shed light on possible marginal spectral
variability in the 2004 observations, we have fit all the spectra
simultaneously with an absorbed blackbody plus power--law model, with
all parameters tied between the spectra. Results are reported in
Tab.\,1 last column, and show that the spectrum does not display
significant variability between the four observations. We also report
in Fig.\,\ref{cntrpl} on the contours plot for the photon index and
absorption parameter for this the joint fit, showing that both
parameters are relatively well constrained by this modelling. However,
looking at the residuals we see a (not significant) excess in the
residuals above 8\,keV (see e.g. Fig.\ref{fig:spectra}), probably due
to the onset of the hard X--ray component (Kuiper et al.~2006;
den~Hartog et al.~2006, 2007). To shed light on this issue, we have
attempted the fit of the March 2004 observation with the addition of
the 25-250\,keV INTEGRAL spectrum (see \S\ref{multiband} for further
details).

We also find some (not significant) deviations from the best fit model
in the lower energy range (see Fig.\,\ref{fig:spectra}) for the two
deepest observations in 2004. However, we believe these deviations
are due to both remaining calibration issues and/or, possibly,
deviations in the ISM abundances from the (assumed) solar ones (see
also Durant \& van~Kerkwijk 2006a).

We have performed a pulse-phase spectroscopy analysis for all the
observations, but we report below only the two deeper observations in
2004, because previous \XMM\, observations gave consistent results but
with a smaller accuracy (see also G\"ohler et al. (2005) for the PPS
of the January 2003 observation).

We have generated 10 phase-resolved spectra for each of the 2004
observations. Phase zero is arbitrarily set close to the start of the
observations (MJD = 53065.38674 and MJD = 53210.30000 for the March
and July observations, respectively) and with all phase bins of the
same size. The choice of the number of intervals has been made a
priori in order to have enough statistics in each phase--resolved
spectra to detect, at a 3$\sigma$ confidence level, a cyclotron line
with an equivalent width $>$ 30 \,eV. This allow us to keep the number
of trials to a minimum in case evidence for a cyclotron line is found.

\begin{figure}
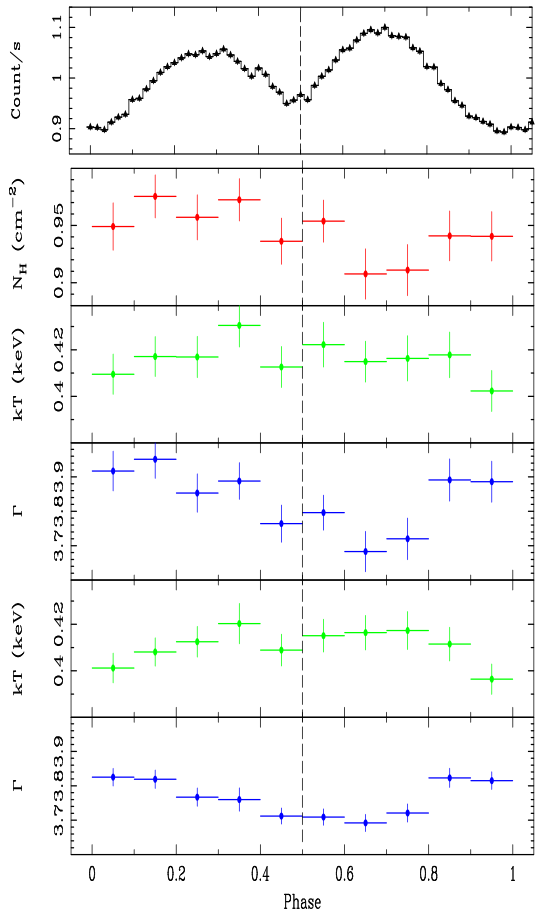

\vbox{
\psfig{figure=efold_pps_final.ps,width=7cm,height=2cm,angle=-90}
\psfig{figure=pps_march_final_c.ps,width=7cm,height=10cm,angle=-90}}
\caption{Top panel reports on the 0.3--10\,keV pulse profile for
    the March 2004 observation. From top to bottom: 2nd, 3rd
    and 4th panels report on the spectral parameters of the pulse
    phase spectroscopy analysis for the \uu\, March 2004 observations
    (see also \S\,\ref{spectra}) with the absorption parameter $N_H$
    free to vary. Last two panels correspond to phase-resolved
    spectral parameters with fixed $N_H =
    1\times10^{22}$cm$^{-2}$. Vertical dashed line is at phase 0.5 for
    clarity.}
\label{fig:phase_paramsspectra}
\end{figure}

The absorbed blackbody plus power law model provides an excellent fits
for all the ten pulse phase-resolved spectra in both observations,
either when leaving \nh\ free (see the 2nd, 3rd and 4th panels from
top in Fig.\,\ref{fig:phase_paramsspectra}) or when fixing it to the
value obtained from the phase-average analysis (see last two panels in
Fig.\,\ref{fig:phase_paramsspectra}).  In all the observations there
is a similar spectral variability with phase, hence we show in
Fig.\,\ref{fig:phase_paramsspectra} only the longer March 2004
observation. The phase-dependent photon index (as obtained from the
fits with fixed absorption, see last panel of
Fig.\,\ref{fig:phase_paramsspectra}) shows deviations from a constant
value with a significance of about 4$\sigma$, while the blackbody
temperature shows a 3.3$\sigma$ variability. The blackbody temperature
seems to vary with phase in anti-correlation with the photon-index, as
also suggested by Patel et al. (2003). Note that G\"ohler et
al.~(2005) did not detect any phase-resolved blackbody temperature
variability because of the much lower number of counts.

However, it should be noticed that the \nh\ value we obtain from this
modelling might be overestimated, since a lower value was found from
fitting the ISM edges directly in the RGS data (Durant \& van~Kerkwijk
2006a) and by using other more physical, although yet less common,
models (see \S\ref{multiband} and Rea et al.~2007a).

No spectral features are detected in the phase-averaged and
phase-resolved spectra of both the pn and RGS data (see also Juett et
al.~2002 for the \CXO\, grating upper limits) of the two 2004 deep
observations. In Tab.\,2 we report the 3$\sigma$ upper limits of the
equivalent width of a Gaussian line with $\sigma_{\rm line}$=0
(narrower than the pn instrumental energy resolution) and $\sigma_{\rm
line}$=100\,eV.  These are the deepest upper limits to date on the
presence of lines in the spectra of an AXP. So far, the only evidence
for an absorption line in an AXP, detected at a significance of
$\sim3\sigma$, was discovered by \BSAX\, (Rea et al.~2003; but see
also Rea et al.~2005, 2007d) in an observation of the AXP \rxj. This
absorption line, tentatively interpreted as due to resonant cyclotron
scattering, was observed at an energy of 8.1\,keV, with a width of
0.2\,keV and an equivalent width (EW) of 0.8\,keV. For \rxj, the
absorption line was observed when the source had a peculiar high
X--ray flux and hard spectrum, while was not detectable during a
subsequent observation (two years later) when \rxj\, was found at a
lower flux level and with a softer X--ray spectrum (Rea et al.~2005,
2007d). Comparing our upper limits on the presence of lines in the
spectrum of \uu\, (Tab.\,2) with the
\rxj\, line properties (Rea et al. 2003), we could have easily
detected in the \uu\, 0.3--9\,keV spectrum any absorption or emission
line with the same width and EW as observed in \rxj.


\begin{table}
\begin{center}
\begin{tabular}{lcc}
\hline
\hline
 \multicolumn{1}{l}{Energy Range} & \multicolumn{1}{c}{$\sigma_{\rm line} =0$ (eV)} &
 \multicolumn{1}{c}{$\sigma_{\rm line} =100$ (eV)} \\
\hline

1--2\,keV & $< 4$ &$< 9$  \\
2--3\,keV &  $< 4$ & $< 7$ \\
3--4\,keV  &  $< 12$ & $< 19$ \\
4--5\,keV  &  $< 16$  &  $< 21$   \\
5--6\,keV  &  $< 25$  &  $< 37$   \\
6--7\,keV  &  $< 46$  &   $< 51$   \\
7--8\,keV  &  $< 88$  &   $< 124 $  \\
8--9\,keV  &  $< 405$  &   $< 416$  \\
9--10\,keV &  $< 527$  &   $< 789$  \\

\hline
\hline
\end{tabular}
\caption{3$\sigma$ upper limits on the equivalent width of a Gaussian 
line with width $\sigma_{\rm line}$=0 (narrower than the pn
instrumental energy resolution) and 100\,eV, derived from the March
2004 \XMM\, spectrum of \uu.}
\end{center}
\end{table}


\section{Multiband spectral modelling}
\label{multiband}

The \uu\, X--ray spectrum has been discovered to have a hard X--ray
component (Kuiper et al.~2006; den~Hartog et al.~2007). Rea et
al. (2007a) applied several models to the 0.5--250\,keV \uu\, spectrum
in order to model the emission of this source across this large energy
range, using simultaneous observations performed in July 2005 with
{\it Swift--XRT} (0.5--8\,keV) and {\it INTEGRAL} (25--250\,keV;
den~Hartog et al.~2007). Here we use the latter {\it INTEGRAL}
spectrum fitting it together with our March 2004 \XMM\, observation
(being the longest one). This multiband fitting is performed in
order to i) asses the presence of an hard component in the 8-10\,keV
\XMM\, spectrum of \uu\, (see Fig.\ref{fig:spectra}), and ii) to test
whether the spectral parameters derived in Rea et al. (2007a), are
consistent with those derived from the higher \XMM\, statistics and
larger energy band (we used here the pn in the 1--10\,keV energy
range).


\begin{table*}
\begin{center}
\begin{tabular}{lrlrlr}
\hline
\hline
\multicolumn{2}{c}{BB + 2PL} & \multicolumn{2}{c}{RCS + PL} & \multicolumn{2}{c}{2 log-parabolae} \\
\hline
$N_{H}$    & $0.926^{+0.005}_{-0.005}$ &  & $0.51^{+0.01}_{-0.01}$      &  & $ 0.53^{+0.01}_{-0.01}$  \\
constant &  1.2   & & 0.86 & &  0.6  \\
kT\,(keV) & $0.422^{+0.001}_{-0.002}$ & kT\,(keV)  &   $0.334^{+0.001}_{-0.002}$   &  $E_{p1}$\,(keV) & $ 1.37^{+0.03}_{-0.04}$  \\
BB\,Flux  &  $ 1.0^{+0.2}_{-0.1}$  &  $\tau_0$ &  $1.84^{+0.07}_{-0.03}$ &    $\beta_1$ & $-3.21 ^{+0.12}_{-0.15}$  \\
$\Gamma_{\rm soft}$      &  $3.86^{+0.01}_{-0.01}$ & $\beta_{th}$ &  $0.22^{+0.01}_{-0.02}$  &  $logP_{1}$~Flux  & $ 2.1^{+1.0}_{-0.9}$ \\
$\Gamma_{\rm hard}$      & $0.8^{+0.1}_{-0.1}$ & $\Gamma_{\rm hard}$ &  $1.1^{+0.1}_{-0.1}$  &  $E_{p2}$\,(keV) & $200^{+80}_{-30}$ \\
PL$_{\rm soft }$\,Flux  &  $7.0^{+1.0}_{-1.0}$  & RCS\,Flux &  $2.2^{+0.3}_{-0.7}$  &   $\beta_2$ & $-0.6^{+0.1}_{-0.1}$  \\
PL$_{\rm hard }$\,Flux  &  $ 1.7^{+0.4}_{-1.2}$ & PL$_{\rm hard}$\,Flux &  $1.5^{+0.8}_{-0.9}$ &  $logP_{2}$~Flux & $ 2.0^{+1.2}_{-1.0}$ \\ \\

Total Abs. Flux     & $2.6^{+0.4}_{-0.5}$  &  &  $3.0^{+0.1}_{-0.4}$ &  & $3.3^{+0.2}_{-1.0}$     \\
Total Flux         & $8.4^{+0.9}_{-0.3}$    &  &  $4.1^{+0.8}_{-0.5}$ &  & $4.4^{+1.1}_{-1.0}$ \\
$\chi_{\nu}^2$ (d.o.f.)     & 1.09 (225) &  & 1.06 (225) & & 0.82 (225)      \\
\hline
\end{tabular}
\caption{Best fit parameters for the 1--250\,keV spectral modelling of
 \uu. Errors are at 90\% confidence level. Fluxes (if not otherwise
 specified) are unabsorbed and in units of
 $10^{-10}$\,erg\,cm$^{-2}$\,s$^{-1}$. RCS, BB, PL$_{\rm soft}$ and
 $logP_{1}$ fluxes are for the 0.5--10\,keV energy band, while PL$_{\rm
 hard}$ and $logP_{2}$ fluxes refer to the 20--250\,keV band.  Total
 fluxes are in the 0.5--250\,keV band. $N_{H}$ is in units of
 $10^{22}$\,cm$^{-2}$. The {\tt constant} parameter, which accounts
 for the inter-calibration, assumes \XMM\, as a reference.}

\end{center}
\end{table*}


We note that, at variance with what has been done in Rea et
al. (2007a), these observations are not simultaneous. However, the
soft X--ray spectrum shows only a very modest variability between the
{\it Swift--XRT} observation performed in July 2005 (simultaneous to
the {\it INTEGRAL} observation we use here) and the \XMM\,
observations we report here.  We then tentatively rely on the
stability of the hard X--ray spectrum and try to fit the overall
spectrum assuming that the source is in the same emission state during
both the \XMM\, and the {\it INTEGRAL} observations.

We fit the 1--250\,keV spectrum with: one absorbed blackbody plus
two power--laws, an absorbed resonant cyclotron scattering model plus
a power--law and two absorbed log--parabolae (Tab.\,3; see Rea et
al. 2007a for further details on these models).

We find that all these models, remove the excess in the $>$8\,keV
residuals (not significant though) found in the \XMM\, spectrum when
fitted with an absorbed blackbody plus a single power--law (see
Fig.\ref{fig:spectra} and Tab.\,1). We then conclude that this excess
in the residuals is indeed due to the presence of the hard X-ray
component. Furthermore, the spectral parameters we find for the
three best fit models are consistent (within 3$\sigma$ statistical
accuracy) with those reported by Rea et al.~(2007a) using the {\it
Swift--XRT} spectrum, with lower statistics and smaller energy
coverage. We note that the constant parameter in the fitting of the
two log--parabolae is slightly smaller than the range of values that
the cross--calibration between \XMM\, and \INT\, should take. This
might be either a hint of the fact that the second log--parabola can
barely take into account the 8--10\,keV part of the spectrum which was
missing in the {\it Swift--XRT} spectrum, or simply a statistical
fluctuation. A simultaneous observation of \uu\, with \XMM\, and a
longer \INT\, exposure might shed light on this issue.

\section{Discussion}
\label{discussion}

We reported here on the timing and spectral analysis of four \XMM\,
observations of the AXP \uu\, performed from 2002 to 2004,
representing the deepest X--ray observations on this source performed
to date.

The pulse profile of 4U 0142+614 is double peaked in the soft X--rays,
and largely variable as a function of energy; we also detect an
increase with energy of the fundamental pulsed fraction component (see
also hints for this increase in Israel et al.~1999; Paul et al.~2000;
G\"ohler et al.~2005). Furthermore, ten years of RXTE monitoring also
revealed the pulse profile to be very variable in time, possibly
related to the presence of glitches (Dib et al.~2007; but see also
Morii et al.~2005).

The presence of two distinct peaks in the phase profile of \uu\,
suggests that the emission might be concentrated in two regions on the
star surface.  Since the two maxima are offset by half a phase cycle,
the emission regions must be located antipodally on the surface of the
star. In a ``standard'' cooling NS, such a geometry of the emission
regions would be naturally produced by a (core-centred) dipolar
magnetic field. In fact, since heat flows preferentially along the
magnetic field lines, in this scenario the polar regions (where the
field lines are more closely packed) turn out to be hotter than the
equatorial belt. On the other hand, for such configurations, the
pulsed fractions are low, the more when gravity effects are accounted
for (e.g. Page 1995; Perna, Heyl \& Hernquist 2000; DeDeo, Psaltis \&
Narayan 2001). The very high pulsed fraction of the hard ($\ga
100$~keV; Kuiper et al.~2006; den~Hartog et al.~2007), photons may be
explained if they are produced in regions high up in the
magnetosphere, where no relativistic effects are expected to wash out
the pulsed emission. This might also tentatively explain the increase
of the pulsed fraction with energy, in particular between the band
dominated by the thermal component, and that dominated by the
power--law. However, much more detailed theoretical models are need to
fully interpret this phenomenon. In fact, it is unlikely that
magnetars are passive coolers in which the surface temperature
distribution is solely determined by the large-scale field
topology. As stressed by Thompson, Lyutikov \& Kulkarni (2002),
portions of the star surface are heated by the returning currents
which flow into a ``twisted'' magnetosphere, and this may be a
possible explanation for the presence of hotter regions on the star
surface. It has been recently suggested (Thompson, Lyutikov \&
Kulkarni 2002; Lyutikov \& Gavriil 2005; Fernandez \& Thompson 2007)
that the mechanism responsible for the formation of the 0.1--10~keV
spectrum of AXPs and SGRs is cyclotron resonant up-scattering of soft
($\sim 0.5$~keV) thermal photons produced at the star surface by
electrons (or pairs) which fill the (twisted) magnetosphere. Currents
are mainly concentrated towards the magnetic equator, and this, as the
Monte Carlo simulations of Fernandez \& Thompson (2007) show, makes
the emission pulsed even if primary thermal photons are produced
uniformly at the star surface. We performed some preliminary
calculations with a Monte Carlo code (Nobili et al. in preparation) in
order to check if the increase of the pulsed fraction with energy is
qualitatively reproduced by the resonant scattering model. For
instance, assuming that the star is an orthogonal rotator seen at an
angle $\sim 90^{\circ}$ with respect to the rotation axis, we find
that the low energy (0.5--2 keV) light-curve should be double peaked
while, at higher energies (7-10 keV), the secondary peak tends to
disappear and the pulsed fraction increases.

At variance with what has been recently found for a few other AXPs
(e.g. \rxj, \ee\, and \ea; Rea et al.~2005; Mereghetti et al.~2004,
and Woods et al.~2004), these very deep X--ray observations of the AXP
\uu\, find the source stable in flux and spectral shape. However,
spectral and flux variability are expected to be connected with the
AXP bursting activity, as e.g. detected for \ea\, (Kaspi et al.~2003;
Woods et al.~2004). Hence, the recent bursting activity of \uu\,
(Kaspi et al.~2007; Gavriil et al.~2007) should have probably
disturbed its steady emission state, and the source might have
recently enhanced its flux and hardened its spectral shape.

Furthermore, we put very strong upper limits on the presence of X--ray
lines in the spectrum of this AXP, either of resonant cyclotron or
atmospheric nature, the deepest upper limits for an AXP to date. This
is a further confirmation that the X-ray spectral lines, possible due
to proton cyclotron scattering, or to absorption by the NS atmosphere,
are either very weak, or not present at all (e.g. suppressed by QCD
processes; Ho \& Lai 2003; van Adelsberg \& Lai 2006) in the spectra
of AXPs, and probably also of SGRs.

 \vspace{1cm}

This paper is based on observations obtained with \XMM, an ESA science
mission with instruments and contributions directly funded by ESA
Member States and the USA (NASA). NR is supported by an NWO
Post-doctoral Fellowship, and thanks Lucien Kuiper for useful
comments, and Vicky Kaspi and Marjorie Gonzalez for useful discussion
about the pulsed fraction variability of this source. SZ thanks PPARC
for financial support. We thank the referee for his/her very useful
suggestions.

\end{document}